\documentclass[trackchanges]{aastex701}

\usepackage{amsmath}
\usepackage{subcaption}

\usepackage{blindtext}
\nolinenumbers

\begin{document}


\title{\boldmath{Modified Cosmology or Modified Galaxy Astrophysics is Driving the z$>$6 JWST Results? CMB Experiments can discover the Origin in the Near Future}}

\author[orcid=0009-0007-4664-4820]{Harsh Mehta}
\email[show]{harsh.mehta@tifr.res.in}  
\affiliation{Department of Astronomy and Astrophysics, Tata Institute of Fundamental Research, 1, Homi Bhabha Road, Mumbai, 400005, India}

\author[orcid=0000-0002-3373-5236]{Suvodip Mukherjee} 
\email[show]{suvodip.mukherjee@tifr.res.in}

\affiliation{Department of Astronomy and Astrophysics, Tata Institute of Fundamental Research, 1, Homi Bhabha Road, Mumbai, 400005, India}

\begin{abstract}

The massive and bright galaxies observed by the James Webb Space Telescope (JWST) at high redshifts ($z > 6$) have challenged our understanding of the Universe. This may require revisiting the physics of galaxy formation and evolution, or modifying the $\Lambda$CDM cosmological model to explain these observations, or both. We show that high-resolution CMB experiments such as the Simons Observatory (or CMB-S4) can measure smoking-gun signatures jointly in weak lensing and kinematic Sunyaev-Zeldovich (kSZ) power spectra, which can shed light on both these scenarios. {An increase in the matter power spectrum at small scales will enhance the number density of dark matter halos at high redshifts, thereby increasing the galaxy formation rate. This will cause enhanced weak lensing signal from these redshifts and also lead to enhanced  patchy-kSZ signal from the epoch of reionization. However, if only galaxy astrophysics is modified, without any modification in the matter power spectrum, then the patchy-kSZ signal gets altered, while the weak lensing signal remains nearly unaltered. We show that we can measure the modified astrophysical and cosmological scenarios at a statistical significance of $10.4\sigma$ (and $29.8\sigma$) from Simons Observatory (and CMB-S4), which will enable a conclusive understanding on what physical process is driving the high-redshift observations of JWST.}
\end{abstract}


\section{Introduction}

The $\Lambda$CDM model has been highly effective in explaining the large-scale structure, the CMB, and the overall evolution of the Universe \citep{1980lssu.book.....P,Dodelson:2003ft,Planck:2015fie,aghanim2020planck}. The advent to study our Universe at smaller scales  has led to the development of high-resolution surveys such as the JWST \citep{sabelhaus2004overview,gardner2006james,pontoppidan2022jwst}.  However, recent JWST observations of surprisingly massive and bright galaxies at high redshifts $z > 10$ challenge the $\Lambda$CDM framework \citep{labbe2023population,baggen2023sizes,harikane2023comprehensive,harikane2025jwst}. 

The $\Lambda$CDM framework prefers hierarchical structure formation, where small structures evolve into larger ones over time \citep{press1974formation,1980lssu.book.....P}. At $z > 10$, the Universe should host only low-mass galaxies—the presumed drivers of cosmic reionization between $z\sim5.2$ and $z\sim10$ \citep{barkana2001beginning,madau2004early,fan2006observational,fan2006constraining,becker2015evidence,bosman2018new,kulkarni2019large,zhu2021chasing,choudhury2021studying}, which grow over cosmic time to form massive galaxies. Yet, JWST has revealed an abundance of bright, evolved galaxies at $z \sim 14$ \citep{labbe2023population,harikane2023comprehensive,harikane2025jwst}, suggesting earlier and more efficient star formation than expected \citep{vogelsberger2020high,zavala2023dusty,lovell2023extreme}. These findings imply that galaxy formation began earlier or proceeded more vigorously than predicted, or that fundamental cosmological or astrophysical assumptions need revision \citep{gupta2023jwst,fakhry2025high,Sokoliuk:2025phe,Chakraborty:2025cbs,Joyce:2014kja,DelPopolo:2016emo,Tulin:2017ara,Bullock:2017xww,Perivolaropoulos:2021jda,Abdalla:2022yfr}.

Modifications in cosmology alter the predicted number of high-$z$ galaxies by changing the dark matter halo abundance, while variations in stellar physics affect the UV luminosities and ionization efficiency, thus shaping reionization history \citep{miralda2000reionization,bullock2000reionization,barkana2001beginning,madau2004early,furlanetto2004growth,morales2010reionization,zaroubi2012epoch,greig2017global}. Reionization leaves imprints on the CMB via Thomson scattering and screening, or by generating secondary anisotropies such as the kinematic Sunyaev–Zel’dovich (kSZ) effect \citep{birkinshaw1999sunyaev,mcquinn2005kinetic,dvorkin2009reconstructing,Hand:2012ui,ACTPol:2015teu,Hill:2016dta,chen2023patchy,Kramer:2025uwv}. The post-reionization kSZ component is constrained to $D_{\ell} = 1.65 \, \mu{\rm K}^2$ at $\ell=3000$ \citep{shaw2012deconstructing,calabrese2014precision}, while the patchy component depends on reionization morphology and the distribution of ionized inter-galactic medium (IGM) surrounding galaxies  across high redshifts \citep{park2013kinetic,paul2021inevitable,chen2023patchy,jain2024probing}. The recent JWST findings suggest that reionization may have begun earlier due to enhanced matter fluctuations, or nonstandard galaxy evolution, or both \citep{robertson2022galaxy,endsley2023jwst,madau2024cosmic,melia2024cosmic,cain2025chasing}.

Enhanced matter fluctuations at high redshifts would increase the number of massive halos and amplify CMB lensing by the evolving gravitational potentials \citep{zaldarriaga1998gravitational,zaldarriaga1999reconstructing,Hu:2000ee,seljak2000lensing,guzik2000lensing}. Future CMB surveys such as the Simons Observatory (SO) \citep{Ade_2019,SimonsObservatory:2018koc} and CMB-S4 \citep{abazajian2016cmbs4} will measure the lensing potential up to $L \sim 2000$, probing small scales ($k > 0.1 \,{\rm Mpc}^{-1}$) at $z>6$, where JWST results are most discrepant with $\Lambda$CDM. Since velocity fields correlate with density fluctuations, these same perturbations also impact the kSZ signal. Therefore, a combined analysis of small-scale kSZ and CMB lensing can disentangle the effects of enhanced matter fluctuations from those of altered galaxy evolution.

An alternative to a modified cosmological explanation is a change in the astrophysics inferred from low-redshift observations. Such a modification would increase emission efficiency and the UV luminosity functions used to estimate high-$z$ galaxy masses \citep{Behroozi:2010rx,Behroozi:2012iw,shibuya2015morphologies,Behroozi:2019kql,Chakraborty:2025cbs}. These functions are also influenced by dust content, bursty star formation, and other non-linear effects \citep{2016MNRAS.460..417S,vogelsberger2020high,zavala2023dusty,sun2023bursty}. A change in stellar emission physics would alter the ionization fraction and optical depth across redshifts. In this work, we do not explore these highly model-dependent aspects. Instead, we test whether a combined study of CMB lensing and kSZ signals can provide insights into the cosmological understanding of structure formation.

With high-resolution CMB data from SO and CMB-S4, we can probe arcminute scales. In this letter, we show how a joint analysis of CMB lensing and kSZ at small scales can address the JWST high-redshift tension within $\Lambda$CDM. We adopt natural units ($\hbar=c=k_B=1$) and Planck 2018 cosmological parameters \citep{aghanim2020planck}.

\section{Effects of Modification to $\Lambda$CDM Cosmology}

JWST observations reveal more massive galaxies at high redshifts than predicted by the $\Lambda$CDM model. Small-scale modifications to $\Lambda$CDM can explain this excess by increasing the abundance of dark matter halos \citep{Joyce:2014kja,DelPopolo:2016emo,Tulin:2017ara,Bullock:2017xww,Perivolaropoulos:2021jda,Abdalla:2022yfr}, implying enhanced matter density fluctuations ($\delta$) at $z>6$. These fluctuations modify the matter power spectrum, defined as
\begin{equation}
\langle \delta(\mathbf{k},z)\delta^*(\mathbf{k'},z) \rangle = (2\pi)^3 \delta^{3}(\mathbf{k - k'}) \,P_{\delta \delta}(k,z),
\end{equation}
where $k = |\mathbf{k}|$.
Large-scale ($k < 0.1\,{\rm Mpc}^{-1}$) power spectra are well constrained by low-redshift surveys such as the Hubble Space Telescope (HST) and Sloan Digital Sky Survey (SDSS) \citep{refregier2002cosmic,gil2015power}, while small-scale behavior depends on non-linear physics. Enhancement in power at small-scales may arise from dark matter clumping, amplified adiabatic perturbations, or cosmic strings \citep{Kolb:1993zz,Germani:2017bcs,Kannike:2017bxn,Garcia-Bellido:2017mdw,jiao2023early,jiao2024n}, whereas neutrino free-streaming, or warm or fuzzy dark matter suppress power \citep{hu1998weighing,lesgourgues2006massive,spergel2000observational,hu2000fuzzy,bode2001halo,viel2005constraining,marsh2016axion}. Baryonic physics such as clustering of cooling stars and Active Galactic Nuclei (AGN)/supernova feedback also affect these scales \citep{semboloni2011quantifying,van2011effects,schaye2015eagle,mead2015accurate}.

{
JWST results suggest enhanced small-scale power and a higher abundance of massive halos \citep{baggen2023sizes,labbe2023population,harikane2023comprehensive,harikane2025jwst}. There are various theoretical scenarios that can lead to enhanced power at small scales \citep{hutsi2023did,fakhry2025high}. In scenarios where axion-like particles make up the dark matter and the associated Peccei–Quinn U(1) symmetry breaks after inflation, the Kibble mechanism induces large-amplitude fluctuations on small spatial scales \citep{kolb1993axion,o2022simulations,ellis2022structure}. The presence of massive primordial black holes (PBHs), either individually or in spatially clustered configurations, leads to shot-noise contributions in the matter power spectrum \citep{carr2018primordial,inman2019early,de2020clustering}. In single-field inflation models, features in the inflaton potential that transiently decelerate the inflaton can lead to enhanced adiabatic power spectra, generally accompanied by a growing curvature power spectrum \citep{garcia2017primordial,kannike2017single,germani2017primordial,karam2023anatomy}. For the first two scenarios, the power-law term in the matter power spectrum corresponds to an isocurvature contribution to matter fluctuations \citep{hutsi2023did,fakhry2025high}. Further transfer function modifications, such as those in relativistic scenarios such as evolution of warm dark matter (WDM),  neutrinos, etc., can also affect the power spectrum \citep{mcnally1995small,viel2012non,binder2016matter,ganjoo2023effects}. Scenarios such as dark matter interactions, modified gravity and early dark energy can affect the transfer function and hence, the power spectrum at small scales \citep{koyama2009non,pettorino2013early,lombriser2015unscreening,binder2016matter,huo2018signatures,sobotka2025signatures}. The COBE/FIRAS bound constrains only the adiabatic component, while constraints on the isocurvature component from $\mu$ and $y$ spectral distortions are much weaker \citep{chluba2013cmb}. Consequently, the cutoff scales in the first two scenarios fall far below the relevant limits. By contrast, for adiabatic power enhancements generated by features in the inflationary potential, COBE/FIRAS constraints rule out PBH formation for cutoff scales $k > 100h \, \rm{Mpc}^{-1}$}, unless the accompanying growth of the curvature power spectrum is sufficiently weak. Modifications from these different scenarios can affect the power spectrum, which can be captured by a scale-dependent and redshift-dependent bias term. Upcoming CMB experiments such as SO and CMB-S4 \citep{Ade_2019,SimonsObservatory:2018koc,abazajian2016cmbs4} will probe small scales ($k>0.1 \,{\rm Mpc}^{-1}$). We parametrize high-$z$ deviations using a phenomenological bias model motivated from generalized bias studies \citep{colin1999evolution,zheng2007breaking,baldauf2011galaxy}:
\begin{equation}
b_{\delta}^2(k,z) = b_{\delta0} + \,b_{\delta z}\,(z/z_0)\,(k/k_0)^2,
\label{eq:bdelta}
\end{equation}
where $b_{\delta0}$ is the fiducial bias, $b_{\delta z}$ quantifies high-$z$ modification, $z_0 = 100$ is degenerate with $k_0$, set at $k_0 = 0.1 \, \rm{Mpc}^{-1}$ being the modification scale.  
The modified non-linear matter power spectrum then becomes
\begin{equation}
P_{\delta \delta}(k,z) = b_{\delta}^2(k,z)\bar{P}_{\delta \delta}(k,z) ,
\label{eq:Pdelta}
\end{equation}
with $\bar{P}_{\delta \delta}$ the standard non-linear $\Lambda$CDM prediction \citep{aghanim2020planck}.
Enhanced small-scale power boosts massive halo formation, strengthens CMB lensing, and also accelerates galaxy evolution and subsequent reionization, amplifying the patchy kSZ signal. In the upcoming sub-sections, we look at the impact of these modifications on CMB lensing and patchy-kSZ signals. 

\subsection{Effect on CMB Lensing}
\label{sec:lensing}
Weak gravitational lensing of CMB photons redistributes temperature and polarization anisotropies across angular scales, determined by the derivative of the line-of-sight integrated lensing potential {\citep{martinez1997effect,zaldarriaga1998gravitational,zaldarriaga1999reconstructing,guzik2000lensing,seljak2000lensing,Hu:2000ee,hu2002mass}}.  The lensing potential depends on the Weyl potential, which depends on the matter density perturbations in the Universe. A modification in the matter power spectrum will change the Weyl potential, and hence the lensing potential \citep{Lewis:2006fu,Hanson:2009kr}.   
 The lensing potential is obtained as:
\begin{equation}
    C_{\ell}^{\psi} = \frac{4\ell^2(\ell + 1)^2}{2\pi}\int_{0}^{\chi_*}\chi d\chi P_{\Psi}(\frac{k_{\ell}}{\chi}  , z(\chi))\left(\frac{\chi_* - \chi}{\chi \chi_*} \right)^2 = \frac{9 \Omega_m^2(z) H^4(z) \ell^2(\ell + 1)^2}{4\pi^3}\int_{0}^{\chi_*}\chi d\chi \frac{P_{\delta \delta}}{k_{\ell}}(\frac{k_{\ell}}{\chi}   , z(\chi))\left(\frac{\chi_* - \chi}{\chi \chi_*} \right)^2, 
    \label{equ:clpsi}  
\end{equation}
where $k_{\ell} = (\ell + 1/2) / \chi$ using Limber approximation,  $\chi_*$ is the comoving distance to the surface of last scattering (at $z \sim 1089$), $P_{\Psi}$ is the Weyl potential, $\Omega_m(z)$ and $H(z)$ are the matter density parameter and Hubble parameter at redshift $z$, and $P_{\delta \delta}$ is the matter power spectrum. 
{Changes in the matter power spectrum at high redshifts will change the potential wells and affect the gravitational interaction of the CMB photons across a broad redshift kernel. Since the gravitational potentials are linked to density perturbations, the Weyl potential is related to the matter power spectrum. We use the relation between the two during matter and dark energy domination. Thus, the modification parameter $b
_{\delta}$ changes the Weyl potential and hence, the lensing potential.} 

\begin{figure}[h!]
    \centering
    \begin{subfigure}[t]{0.45\textwidth}
\includegraphics[width=9.2cm]{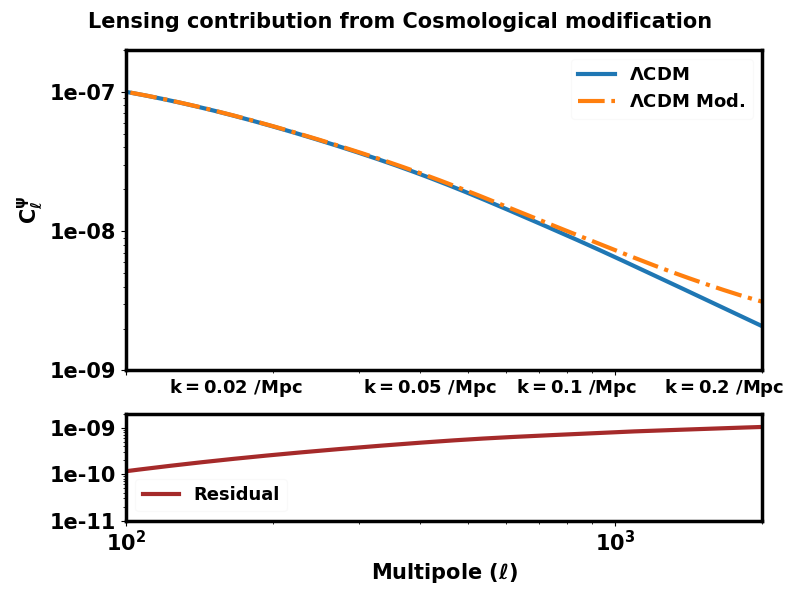}
    \caption{Impact on Lensing Potential power spectrum.}
    \label{fig:lensk}
    \end{subfigure}
    \hspace{1cm}
    \raisebox{0.1cm}{
    \begin{subfigure}[t]{0.45\textwidth}
\includegraphics[width=9cm]{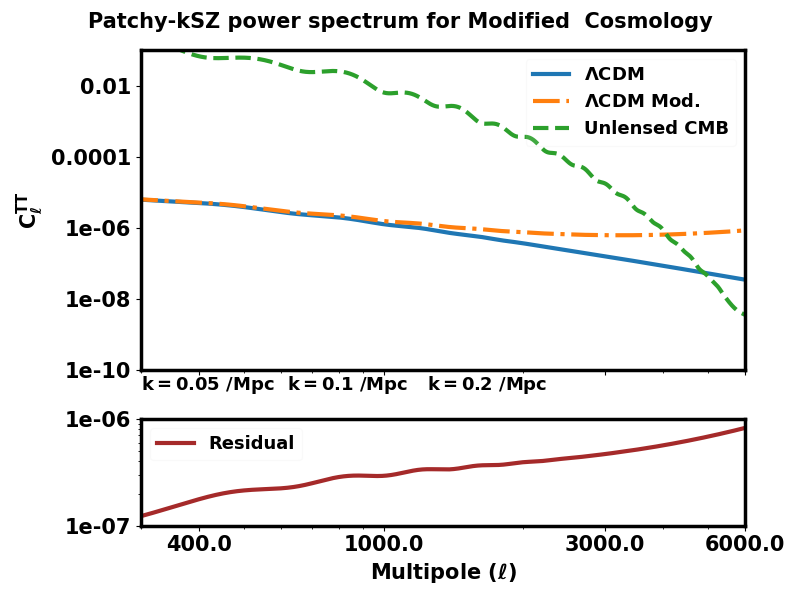}
    \caption{Impact on patchy-kSZ power spectrum.}
    \label{fig:kszk}
\end{subfigure}
}
\caption{The integrated lensing potentials $C_{\ell}^{\psi}$ (\ref{fig:lensk}) and patchy-kSZ power spectra  (\ref{fig:kszk}) for the cases of $\Lambda$CDM (blue solid) and modified $\Lambda$CDM (orange dot-dashed). The deviation between $\Lambda$CDM and modified cosmology increases at higher multipoles $\ell > 800$ as can be seen in the residual plot in the lower panel. The detectors SO and CMB-S4 will be able to estimate the lensing and patchy-kSZ power spectra at these high multipoles, which can be used to probe small-scale modifications in $\Lambda$CDM, as well as the underlying astrophysics which affects the electron density bias.  The high redshifts ($z > 6$) contribute to these multipoles at scales $k > 0.1 \, \rm{Mpc^{-1}}$, where JWST shows an excess of massive galaxies. Also plotted in \ref{fig:kszk} is the unlensed CMB power spectrum (green-dashed) which dominates in power up to multipoles $\ell >3000 $. The patchy-kSZ signal will also increase or decrease if the electron density bias $b_e$ is higher or lower. }
\label{fig:bothplots}
\end{figure}

{Lensing causes mode-mixing of multipoles ($\ell$) with each other separated by $L$, with the mixing depending on the power at the lensing multipole ($L$). With current and upcoming detectors such as SO and CMB-S4, we will be able to measure the lensing potential up to high multipoles of $L \sim 1500$. This high resolution will allow us to probe the lensing contribution from high redshifts as shown in Fig. \ref{fig:lensk}.  The small scales of $k > 0.1 \, \rm{Mpc^{-1}}$ are not very well studied due to resolution and sensitivity constraints, and the impact from non-linear dynamics. Using the Limber approximation ($k \chi(z)  \sim L + 1/2$), the scales $k > 0.1 \, \rm{Mpc}^{-1}$ at high redshifts $z > 6$ contribute at multipoles $L > 800$. If there are a higher number of massive halos at high redshifts than expected from $\Lambda$CDM, these will lead to enhanced lensing from these redshifts. The modification scenario is shown in orange dot-dashed line compared with the blue solid line for $\Lambda$CDM case. The increase in lensing potential is evident at high multipoles due to enhanced lensing from high redshifts. Thus, any modification from $\Lambda$CDM scenario can be probed using multipoles $L > 800$.} 

{Any modification in $\Lambda$CDM matter power spectrum will inevitably impact the lensing potential, as it is integrated over a broad redshift kernel. If the modification in power spectrum is different from the $k^2$ dependence assumed in this analysis, it would affect the lensing potential at different multipole ranges. Thus, CMB lensing provides a robust probe of small-scale modifications to $\Lambda$CDM.

\subsection{Effect on patchy-kSZ}
\label{sec:kszonly}

 {The ionization of the intergalactic medium (IGM) around the galaxies due to stellar emission leads to the production of free electrons within the ionized bubbles \citep{barkana2001beginning,madau2004early,fan2006observational,fan2006constraining,becker2015evidence,bosman2018new,kulkarni2019large,choudhury2021studying,zhu2021chasing,chen2023patchy}. 
 These bubbles increase in size as the number of ionizations, and hence the ionization fraction increases with time. The Doppler-shifting of CMB photons that scattered off these free electrons with bulk velocities leads to the kSZ effect in the CMB \citep{birkinshaw1999sunyaev,mcquinn2005kinetic,Hand:2012ui,ACTPol:2015teu,adam2016planck,Hill:2016dta}. During the epoch of reionization, the Universe is partially ionized with the kSZ effect depending on the duration and  timing  of this epoch, as well as the morphology of these ionized regions \citep{mcquinn2005kinetic,dvorkin2009reconstructing,park2013kinetic,park2016impact,paul2021inevitable,chen2023patchy,jain2024probing}. This leads to the patchy-kSZ signal, as opposed to the homogeneous kSZ signal, which is a result of the scattering of CMB photons after reionization of the Universe \citep{shaw2012deconstructing,calabrese2014precision,park2016impact,chen2023patchy}.  
 
The kSZ effect depends on both the free electron density and peculiar velocity fields. The free electron density field depends on the galaxy formation and evolution, as well as the efficiency of the galaxies to ionize the intergalactic medium \citep{gnedin2000effect,bullock2000reionization,barkana2001beginning}. The peculiar velocity fields depend on  the gravitational potential, which  depends on the matter density fluctuations in the Universe \citep{1980lssu.book.....P,Dodelson:2003ft,aragon2013hierarchical}. The kSZ power spectrum is obtained as :
\begin{equation}
    C_{\ell}^{\rm{kSZ}} = (\sigma_T \bar{n_{e0}}T_{0})^2 \int d\chi \frac{\exp{[-2\tau(\chi)]}}{2a^{4}\chi^2} P_{q_{\perp}}(k = (\ell + 1/2)/\chi,\chi),
\label{eq:kszspec}
\end{equation}
where $\sigma_T$ is the Thomson scattering cross-section, $\bar{n_{e0}}$ is the mean electron number density, $T_0$ is the CMB temperature, $\tau(\chi)$ is the optical depth to the comoving distance $\chi$, and $P_{q_{\perp}}(k,z)$ is the transverse momentum power spectrum, using the Limber approximation ($k\chi = \ell + 1/2$), which is obtained as the power spectrum of the transverse component of momentum $\mathbf{q}$, given as:
\begin{equation}
    \mathbf{q(x)} = [1 + \delta_e(\mathbf{x})]\mathbf{v}_e(\mathbf{x}),
\end{equation}
{where $\delta_e$ are the  electron density fluctuations about the mean ($\delta_e(\mathbf{x}) = (x_e(\mathbf{x}) - \bar{x_e}) / \bar{x_e}$), with $\bar{x_e}$ being the mean ionization fraction at the corresponding redshift, and $x_e$ is defined as  $x_e$ = $n_e$/$n_H$ , where $n_e$ and $n_H$ are the mean electron and Hydrogen (neutral or ionized) number densities,} and $v_e$ are their peculiar bulk velocities. 
The momentum $\mathbf{q}$ depends on both the electron density, and velocity fields, both of which depend on the underlying matter distribution. A modification to the underlying dark matter field will change the velocity field, as well as the electron density field by providing potentials for the formation of galaxies. Along with this, the production rate of ionizing photons from galaxies as a function of cosmic time will impact the momentum field.  The evolution of the ionization fraction ($x_e$) can then be written as \citep{madau1999radiative,barkana2001beginning,furlanetto2006cosmology}:
\begin{equation}
    \frac{dx_e}{dt} = \frac{dN_{\rm{ion}}}{dt} \frac{1}{n_H} -\frac{x_e}{t_{\rm rec}},
    \label{eq:xe}
\end{equation}
where $n_H$ is the number density of Hydrogen, $d N_{\rm{ion}} / dt$ is the rate of the ionizing photons density being produced, and the second term accounts for the number of recombinations that hydrogen atoms undergo in the time-scale of $t_{\rm rec}$. The rate of hydrogen-ionizing photons is  related to the UV luminosity function as
\citep{madau1998star,robertson2013new,bouwens2015uv}.

\begin{equation}
    dN_{\rm{ion}} / dt = \int_{-\infty}^{M_{UV}^*} \Phi(M_{UV})L(M_{UV})\xi_{\rm{ion}}dM_{UV} f_{\rm{esc, UV}},
    \label{eq:dniondt}
\end{equation}
where $f_{\rm{esc, UV}}$ is the escape fraction of photons, which can depend on the UV flux\citep{madau1998star,robertson2013new,bouwens2015uv}, $\xi_{\rm{ion}}$ is the ionization efficiency in terms of the rate of hydrogen-ionizing photons per unit UV luminosity, $L(M_{\rm UV})$ is the luminosity -magnitude relation, and $\Phi(M_{\rm UV})$ being the UV luminosity function, generally modelled by a Schechter function \citep{schechter1976analytic}. {
$M_{UV}^*$ marks the limiting magnitude at the low luminosity end, with typical values ranging between -10 and -17 depending on the dataset being considered. }

}

{The electron density field will also vary based on the redshift being considered and the morphology of the ionized bubbles \citep{furlanetto2005taxing,iliev2006simulating,zahn2007simulations,friedrich2011topology,shukla2016effects,furlanetto2016reionization,chardin2018self,chen2019stages,gazagnes2021inferring}. The patchy-kSZ is a result of the integrated Doppler-shifted scattering of CMB photons due to these ionized regions of different sizes across different redshifts. This scattering will depend on the size of the bubbles and redshift. Thus, the patchy-kSZ power spectrum at different multipoles ($\ell$) will get contributions from a range of redshifts, depending on the sizes of the bubbles being considered.  The Limber approximation ($k\chi(z) \approx \ell + 1/2$)  takes into account the varying scattering across these redshifts ($z$) due to different bubble morphologies, defined by the scale $k$.  Hence, there will be an overall redshift-integrated electron density bias $b_e$ involved which will account for the electron density fluctuations based on the evolution of galaxies in the Universe, as well as their ionizing efficiencies \citep{paul2021inevitable}. The electron density bias will depend on the overall number of ionizations, and  will be higher for higher ionization efficiencies, as there will be more electrons that will scatter the CMB photons. 
   Thus, we model the change in patchy-kSZ power spectrum using the modified transverse momentum power spectrum as: }\begin{equation}
    P_{q_{\perp}}(k,z) =  b_{\delta}^2(k,z) b_{e}^2(k,z) P_{\delta \delta}(k,z) .
    \label{eq:pqsum}
\end{equation}  

{

 We show the patchy-kSZ signal for the cases of $\Lambda$CDM (blue solid) and modified-$\Lambda$CDM (orange dot-dashed) in Fig. \ref{fig:kszk}. Using Limber approximation ($k \chi(z)  \sim \ell + 1/2$), the scales $k > 0.1 \, \rm{Mpc}^{-1}$ at high  redshifts $z > 6$ contribute at multipoles $\ell > 800$. If there are a higher number of massive halos at high redshifts than expected from $\Lambda$CDM, these will lead to enhanced matter density fluctuations, which will increase the peculiar velocity and electron density fluctuations at these redshifts. The modification scenario is shown in orange dot-dashed line, against the blue solid line for unmodified $\Lambda$CDM case. The increase in patchy-kSZ is evident at high multipoles due to enhanced velocity and electron density fluctuations. Also, the CMB power spectrum (green-dashed) is shown which dominates at low multipoles ($\ell < 3000$). Thus, any modification from $\Lambda$CDM scenario can be probed using multipoles $\ell > 3000$, requiring measurement of the patchy-kSZ signal at higher multipoles. With the current and upcoming detectors such as SO and CMB-S4, we will be able to measure the temperature power spectrum up to high multipoles of $\ell \sim 6000$. Thus, they will play a crucial role in being able to probe any such small-scale modifications in $\Lambda$CDM. Since the patchy-kSZ signal at different multipoles will depend on the redshifts and sizes of the ionized bubbles involved in scattering, probing the electron density bias at high multipoles will provide insight into the evolution of these ionized bubbles and the underlying galactic systems across different redshifts \citep{chen2023patchy,jain2024probing}. } { Since halo formation is a non-linear effect, we have included the non-linearity in the matter power spectrum $P_{\delta \delta}$ . The linear electron-density bias assumption is valid on scales greater than about a few Mpc when the galaxy properties are determined by the local environment. Since ionization is determined in the IGM surrounding the galaxies, this is a reasonable assumption,  given the measurement uncertainties at very small scales (large $\ell$).}

{Any modification in $\Lambda$CDM matter power spectrum will impact the patchy-kSZ power spectrum, as it will affect both the fluctuations in electron density, as well as the velocity fields. If the modification in power spectrum is different from the $k^2$ dependence assumed in this analysis, it would affect the patchy-kSZ at different multipole ranges. Varying astrophysical effects, such as the amount of baryons involved in star formation, accompanied with the escape fraction of ionized photons, etc., would also impact the patchy-kSZ power spectrum. Measuring patchy-kSZ up to higher multipoles can probe various cosmological as well as astrophysical phenomena \citep{Kolb:1993zz,spergel2000observational,marsh2016axion,hu2000fuzzy,bode2001halo,viel2005constraining,semboloni2011quantifying,van2011effects,schaye2015eagle,mead2015accurate,Garcia-Bellido:2017mdw,Germani:2017bcs,Kannike:2017bxn,tulin2018dark}.
}
In the next section, we show how these high redshift JWST observations can be explained using modifications in $\Lambda$CDM cosmology.

\section{OBTAINING THE JWST BEST-FIT COSMOLOGICAL MODEL}
\label{sec:jwst}

{The matter field can be realized in the halo model as being composed of dark matter halos which form from gravitational collapse, independent of any stellar astrophysics \citep{1980lssu.book.....P,cooray2002halo,smith2003stable}. The number of these halos can be estimated using the Tinker halo mass function $dn/dM_h$ \citep{tinker2008toward} (calculated using Cluster Toolkit \citep{2022ascl.soft09004M}). The Tinker mass function is an improvement over Press-Schechter \citep{press1974formation} and Sheth-Tormen \citep{sheth1999large} mass functions, as it is obtained from a large suite of high-resolution N-body simulations over a wide range of halo masses, redshifts and overdensity definitions. The general form of a halo mass function is defined as:
\begin{equation}
dn/dM_h = f(\sigma)\frac{\rho_c \Omega_m}{M_h}\frac{d \,\ln \sigma^{-1}}{dM_h},
\label{eq:dndmeq}
\end{equation}
where $\rho_c$ is the critical density of the Universe and $\Omega_m$ is the matter density parameter.
Here $\sigma$ is the root-mean-square (RMS) deviation for a top-hat filter-smoothed ($W$) initial density fluctuation field, which is given as:
\begin{equation}
    \sigma^2(M) = \int_{0}^{\infty} \frac{k^2dk}{2\pi^2} P_{\delta \delta}(k)|W(kR)|^2,
\end{equation}
where $M$ is related to $R$ via, $M = \frac{4\pi}{3}\bar{\rho_m}R^3$, with $\bar{\rho_m}$ being the mean energy density of matter in the Universe. The  $f(\sigma)$ refers to  the halo-multiplicity function, which in the case of Tinker mass function is calculated as \citep{tinker2008toward}:.\begin{equation}
f(\sigma) = B\left[ \left(\frac{\sigma}{e}  \right)^{-d} + \sigma^{-f}  \right] \exp(-g/\sigma^2),
\label{eq:multiplicity}
\end{equation}
with the values of the parameters being $d = 1.97, \, e=1.00, \, f =0.51, \, g = 1.228$, and the normalization factor $B = 0.482$.}

{Any modifications in the matter power spectrum $P_{\delta \delta}(k,z)$ will modify the multiplicity function $f(\sigma)$. Any enhancement in power at small scales will change the relative number of high to low mass halos. This will change the shape of the halo mass function $dn/dM_h$, by changing the number of high mass halos with respect to the low mass halos.    
Hence, the halo mass function 
will depend on the power spectrum $P_{\delta 
\delta}(k,z)$ at different scales and different redshifts.} Higher values of $dn/dM_h$ at high redshifts would increase the number of halos, leading to higher lensing potential at small scales. 
These dark matter halos host galaxies, the formation and evolution of which involve highly uncertain astrophysics in the form of stellar mass-halo mass relations and the modelling of UV luminosities, which are used to infer the stellar masses \citep{Behroozi:2012iw,shibuya2015morphologies,Behroozi:2019kql}. {This involves high uncertainty in the modelling of radiative processes and star formation dynamics \citep{Behroozi:2010rx,vogelsberger2020high,zavala2023dusty,Chakraborty:2025cbs}.

The stellar luminosity of galaxies depends on the astrophysics of galactic formation and evolution, the mass-to-light ratio, as well as the intrinsic mass-luminosity relationship. These suffer from high uncertainties depending on the systems and the redshifts being considered. We employ a simplistic model, whereby 
the stellar absolute magnitudes ($M_{UV}$) can be linked to the halo masses ($M_h$), using the power-law stellar luminosity-halo mass dependence $L_{UV} \propto M_{h}^{\gamma}$ via the relation \citep{Behroozi:2010rx,Behroozi:2012iw,shibuya2015morphologies,Behroozi:2019kql}: 
\begin{equation}
     M_{UV} = a - B(\gamma)\log_{10}{(\epsilon f_b M_h) \equiv A - B\log_{10}{M_h},
}
\label{eq:abparams}
 \end{equation}
where $f_b$ is the cosmic fraction of matter in the form of baryons $\sim 0.156$ \citep{aghanim2020planck} and $\epsilon$ is the fraction of baryons involved in stellar formation, with maximum value being 1, and $a$ and $B$ (dependent on $\gamma$) are the fitting parameters, which vary with redshift and the systems being observed. This modelling would give us three parameters ($\epsilon$, $a$ and $\gamma$) which can change the shape of the luminosity function based on the underlying astrophysics of galactic formation and evolution. But these parameters are correlated, being dependent on each other, as the UV luminosities will depend on the amount of baryons involved in star formation, as well as the escape fraction of photons (see Eq. \eqref{eq:dniondt}).  
To relate the halo masses with the stellar magnitudes, we can use a complex model taking into account all these: the baryon fraction involved in galaxy formation, the mass-to-light ratio, etc. Since these quantities are highly uncertain at high redshifts, we reduce the degrees of freedom using a simple redshift-dependent halo mass-stellar magnitude two parameter relation, that converts halo mass to absolute UV magnitude $M_{UV}$ using parameters $A$ and $B$.}

{We can then obtain the UV luminosity function using abundance matching \citep{vogelsberger2020high,lovell2023extreme,zavala2023dusty,Chakraborty:2025cbs} as:

\begin{equation}
 \Phi =  (dn  / d M_h)|dM_h  / d M_{UV}|  = f(\sigma)\frac{\rho_c \Omega_m}{B}\frac{d \,\ln \sigma^{-1}}{dM_h}
.
\end{equation}
This gives us our phenomenological model for luminosity function in case of modifications in $\Lambda$CDM cosmology.

The luminosity function is modelled using Schechter function \citep{schechter1976analytic}, which is characterized by a power law variation of luminosity with magnitude at the faint end, while an exponential suppression of bright objects. 
 JWST predicts an excess of these bright galaxies at high redshifts than expected from $\Lambda$CDM, thus requiring a change in the luminosity function shape to account for this excess. }  
Using these equations, we fit our phenomenological model at different redshifts to match the JWST Schechter function fits from \citet{harikane2025jwst}. 
and show how a modification to the power spectrum can explain the observed excess of bright galaxies. We use the fiducial value of $b_{\delta 0} = 1$, so that the matter power spectrum at large scales is the one expected from $\Lambda$CDM. The parameters  $A$ and $B$ allow us to change the astrophysics connected with galaxy formation and evolution. Also, we introduce modifications to the power spectrum using the parameter $b_{\delta z}$. 

\begin{figure}
    \centering    \includegraphics[width=18.3cm, height=6cm]{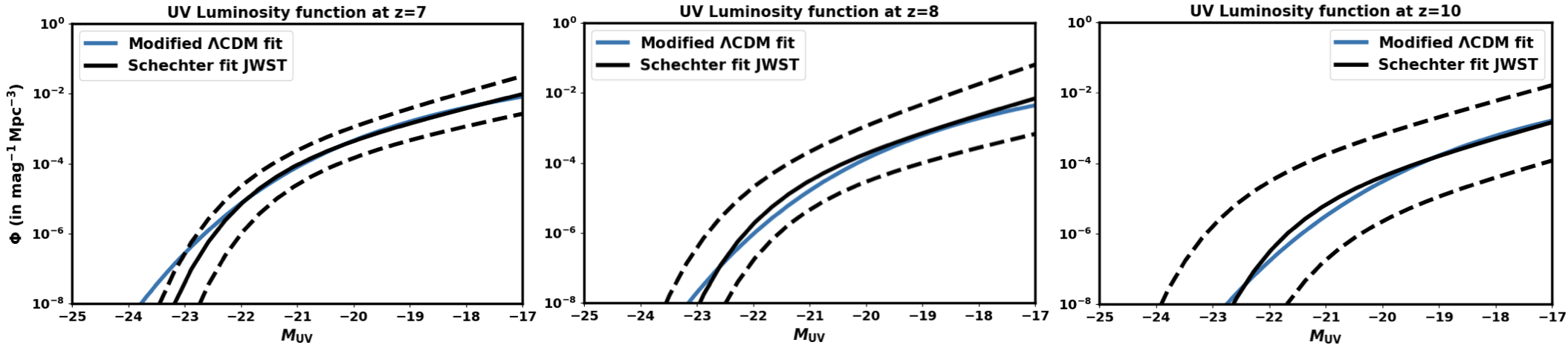}
    \caption{The fits obtained using our phenomenological model at redshifts 7, 8 and 10. Also plotted are the Schechter fits from \citet{harikane2025jwst} with $1\sigma$ variations on them (dashed).}
    \label{fig:dndm}
\end{figure}

\begin{table}[h!]
\centering

\begin{tabular}{|c|c|c|}

\hline

\textit{\textbf{Fitting parameters}} &
\textit{\textbf{Allowed Range}}&
\textit{\textbf{Fit Values}}
\tabularnewline \hline
$b_{\delta z}$ & 0 to 5 & $1.15 \pm 0.87$ \\
 \hline
$A $ & 30 to 50 & $32.3 \pm 1.01$, $36.9 \pm 1.13$
, $48.3 \pm 1.49$
\\
 \hline
$B $ & 1 to 6 & $4.23 \pm 1.21$, $4.57 \pm 1.37$, $5.48 \pm 2.96$ 
\\
\hline

\end{tabular}
\caption{Fitting parameters obtained for our phenomenological model by fitting at the redshifts z = 7, 8, and 10.}
\label{tab:fitpar}
\end{table}

We fit our phenomenological model to the Schechter fit from JWST observed galaxies in \citet{harikane2025jwst} at redshifts 7, 8 and 10. 
The range of parameters within which they are allowed to vary is given in Table \ref{tab:fitpar}. We vary six parameters (3 $A$'s and 3 $B$'s, corresponding to each redshift $z = 7,8$ and 10) and an additional parameter in $b_{\delta z}$ (common to all redshifts) using Scipy's curvefit \citep{2020SciPy-NMeth}. The parameter values so obtained are shown in Table \ref{tab:fitpar}. The phenomenological fit and the Schechter fit from  \citet{harikane2025jwst} are shown in Figure \ref{fig:dndm}, along with the $1\sigma$ variation on the Schechter fit. We use the $1\sigma$ values to calculate the reduced-$\chi^2$ for these three epochs.  We find a $\chi^2_{\rm{red}} = 1.05, 1.04, 1.06 $ for our phenomenological fits at $z = 7, 8$ and $10$ respectively.

This shows that a modification to the power spectrum can explain the high redshift observations of bright and massive galaxies by JWST. This points to a higher number of dark matter halos at high redshifts, which would challenge our understanding of structure formation and opens the window for small-scale phenomenological studies \citep{Joyce:2014kja,DelPopolo:2016emo,Tulin:2017ara,Bullock:2017xww,Sailer:2021yzm,Perivolaropoulos:2021jda,Abdalla:2022yfr,menci2022high,lovell2023extreme,maio2023jwst,Dolgov:2023ijt,huang2024high}. {If the modification in power spectrum is different from the $k^2$ dependence assumed in this analysis, it would affect the shape of the halo mass function ($dn / dM_h$), as well as the luminosity function ($\Phi$). But this formalism of connecting phenomenological models with high redshift observations would allow us to determine if any such variations are possible.}
 We next turn to the aspect of probing any modifications in $\Lambda$CDM cosmology with a joint analysis of CMB lensing and patchy-kSZ signals.

\section{Joint Analysis using CMB Lensing and Patchy-kSZ}

\begin{figure}
\centering
\includegraphics[height=11cm,width=19cm]{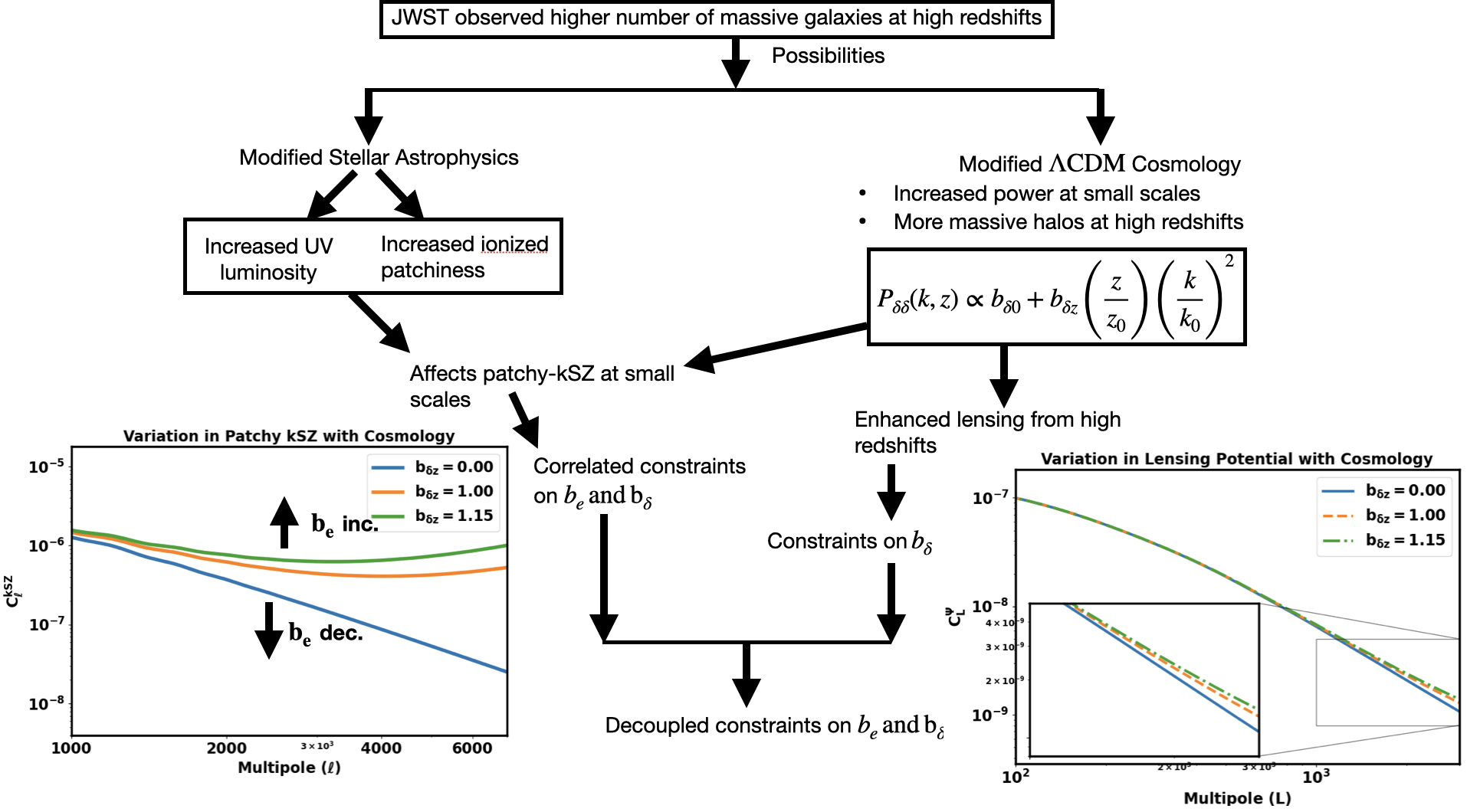}
\caption{JWST-observed massive galaxies can be explained by modified stellar astrophysics (increased ionized patchiness or UV luminosities) or modified $\Lambda$CDM cosmology (increased small-scale power enhancing halo abundance). Both affect the patchy-kSZ signal, but CMB lensing—sensitive only to the matter distribution—constrains $b_{\delta}$, allowing separation of cosmological ($b_{\delta}$) and astrophysical ($b_e$) effects.}
\label{fig:flow}
\end{figure}

The patchy-kSZ contribution to the CMB power spectrum depends on both the electron density bias $b_e$ and the dark matter power spectrum modification $b_{\delta}$, since it arises from galaxy evolution, ionization, and underlying matter fluctuations that shape electron density and velocity fields. The lensing potential depends only on $b_{\delta}$, tracing the total matter distribution. The parameter $b_e$ represents the integrated astrophysical contribution to the patchy-kSZ signal across reionization \citep{paul2021inevitable}. Its variation across multipoles reflects the patchiness in the ionization fraction $x_e$, which increases with the ionization efficiency and the UV luminosity function $\Phi(M_{UV})$ (see Sec. \ref{sec:jwst}). Greater patchiness and higher UV luminosities lead to larger $b_e$ values.

\begin{table}
\centering
\begin{tabular}{|c|c|c|}
\hline \textit{\textbf{Modification scenarios}} & \textit{\textbf{Role in Patchy-kSZ}}&\textit{\textbf{Role in Lensing}} \\
\hline
\hline
Modified Stellar Astrophysics & Measures high $b_{e}$. & No significant effect on lensing signal. \\
\hline
Modified Cosmology & Measures high $b_{e}$. & Measures high $b_{\delta_0}$ and $b_{\delta_z}$. \\
\hline
\end{tabular}
\caption{The role of modifications on patchy-kSZ and CMB lensing signals, and the parameters that can be inferred from
patchy-kSZ and CMB lensing.}
\label{tab:kszlensall}
\end{table}

A joint analysis of the kSZ and CMB lensing signals can separate the effects of $b_{\delta}$ and $b_e$ (see Table \ref{tab:kszlensall}), distinguishing between cosmological and astrophysical origins of the patchy-kSZ signal as explained in Fig. \ref{fig:flow}. The higher number of bright galaxies can be explained either
by modifying stellar astrophysics at high redshifts, or by modifying fiducial $\Lambda$CDM cosmology. Both
can account for JWST observations, and would affect the patchy-kSZ signal. These
effects are parameterized using $b_e$ and $b_{\delta}$ respectively, but cannot be disentangled.  CMB lensing, sensitive only to $b_{\delta}$, can break their degeneracy. 

The JWST best-fit modified $\Lambda$CDM parameters  generate the input lensing and patchy-kSZ power spectra (Eqs. \eqref{equ:clpsi} \& \eqref{eq:kszspec}). A change in the lensing potential redistributes CMB temperature and polarization, suppressing TT and EE peaks while enhancing small-scale power and inducing B-modes \citep{Lewis:2006fu,Hanson:2009kr,Maniyar:2021msb,jain2023framework}. The all-sky formalism of \citet{Hu:2000ee} is used to compute lensed TT, EE, BB, and EB spectra.
Modifications to the matter power spectrum affect the high-redshift halo abundance, enhancing the lensing potential at $\ell > 800$ and increasing the patchy-kSZ signal through enhanced galaxy formation and subsequent IGM ionization \citep{park2013kinetic,chen2023patchy,jain2024probing}. We adopt $\tau = 0.055$ and a tanh ionization profile \citep{aghanim2020planck}, with  $b_e^2 = 4\times10^{-7}$ \citep{paul2021inevitable} at all multipoles, which in general is model-dependent. The CMB power spectra are generated using CAMB \citep{2011ascl.soft02026L}, including foregrounds (synchrotron, dust, free-free, tSZ and CIB) from PySM \citep{Thorne_2017} and masking 50$\%$ of the Galactic plane. Instrumental beams and noises for SO and CMB-S4 frequencies are used \citet{Ade_2019,SimonsObservatory:2018koc,abazajian2016cmbs4}. Post- and patchy-reionization contributions are included, adopting a fiducial $D_{\ell}=1.65\,\mu$K$^2$ for the post-reionization contribution \citep{shaw2012deconstructing,calabrese2014precision}. {The post-reionization kSZ signal is degenerate with patchy-kSZ. The post-reionization power spectrum though is expected to have a different shape  as compared to the patchy-kSZ spectrum, with the patchy kSZ spectrum depending on the ionized bubble sizes at various redshifts, and the post-reionization component being a flat spectrum over the relevant multipoles \citep{shaw2012deconstructing,calabrese2014precision,Park:2011un}. The two spectra can be separated using power spectrum observations over a wide range of multipoles in a joint analysis \citep{choudhury2021cosmic,jain2024probing}. Also, kSZ bispectrum tomography can be used, which  isolates the redshift dependence of the kSZ signal by cross-correlating small-scale CMB temperature fluctuations with tomographic large-scale-structure bins, allowing the measured bispectrum to map electron momentum as a function of redshift. Because post-reionization kSZ arises mostly from halos and filaments at low redshifts, while patchy kSZ originates from ionization bubbles at $z > 6$, their contributions peak in distinct tomographic bins and have different small-scale responses, which can then be used to separate them. CMB-S4, along with DESI will be able to measure the post reionization kSZ with a SNR of around 500 using kSZ tomography  \citep{smith2016detecting,Smith:2018bpn,sato2021kinetic,schaan2016evidence,baxter2019constraining,schaan2021atacama,padmanabhan2023alleviating,raghunathan2024first,sabti2024insights}. 
}

 Using these spectra, we compute the covariance for the minimum-variance lensing \citep{Okamoto:2003zw} and quadratic patchy-kSZ estimators \citep{Dodelson:2003ft,aghanim2020planck}, incorporating Gaussian noise. These JWST-motivated spectra are then used for joint parameter estimation of $b_{\delta}$ and $b_e$ using CMB lensing and patchy-kSZ.

\section{Results}
\begin{figure}
    \centering    \includegraphics[width=16cm, height=8cm]{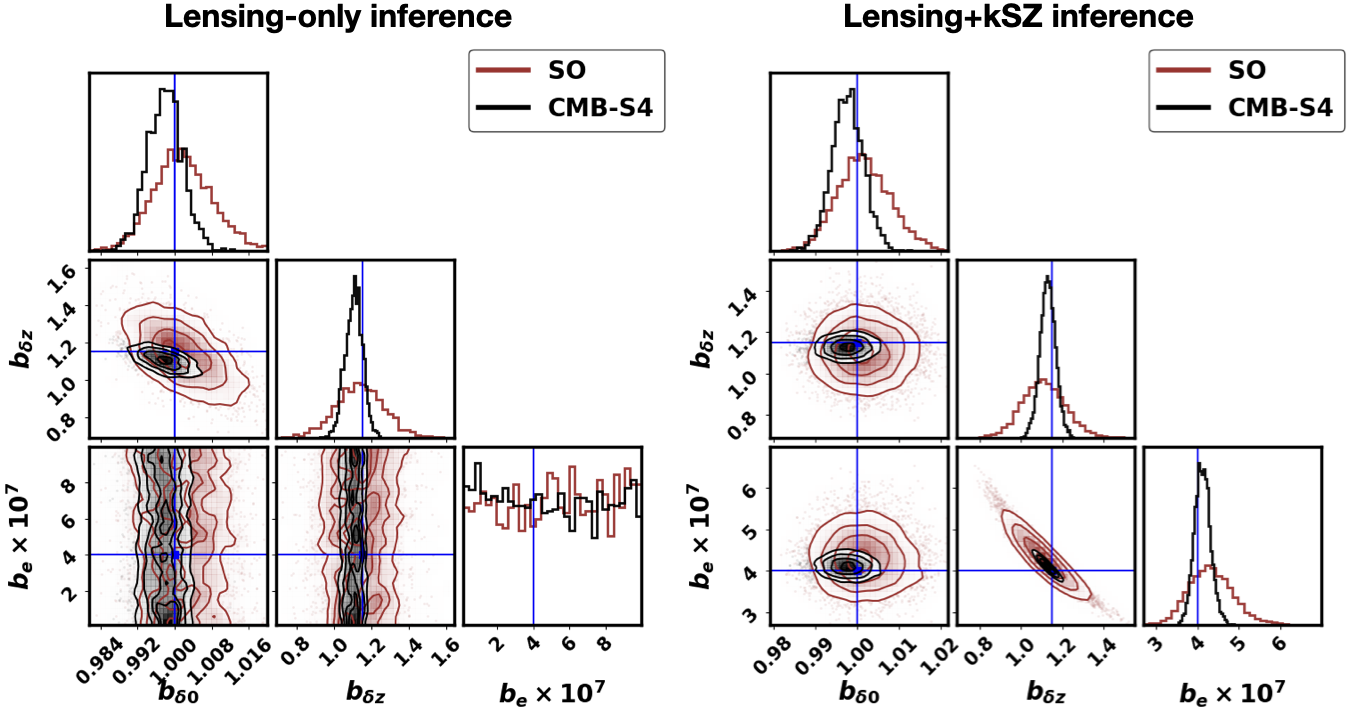}
    \caption{The constraints obtained on modification parameter  ($b_{\delta z}$), and electron density bias ($b_e$) from SO and CMB-S4. {We show the inference for the case of only-lensing estimation (left) and for lensing + kSZ estimation (right).}  The parameters $b_{\delta z}$ and $b_e$ are anti-correlated and can be decoupled using both CMB-lensing and patchy-kSZ observations. The blue lines represent true injected values for JWST best-fit cosmology.{Lensing is able to constrain cosmological modifications, while kSZ can constrain modifications in the understanding of galactic  astrophysics as well.}}
    \label{fig:bayes}
\end{figure}

\begin{figure}
    \centering    \includegraphics[width=14cm, height=9cm]{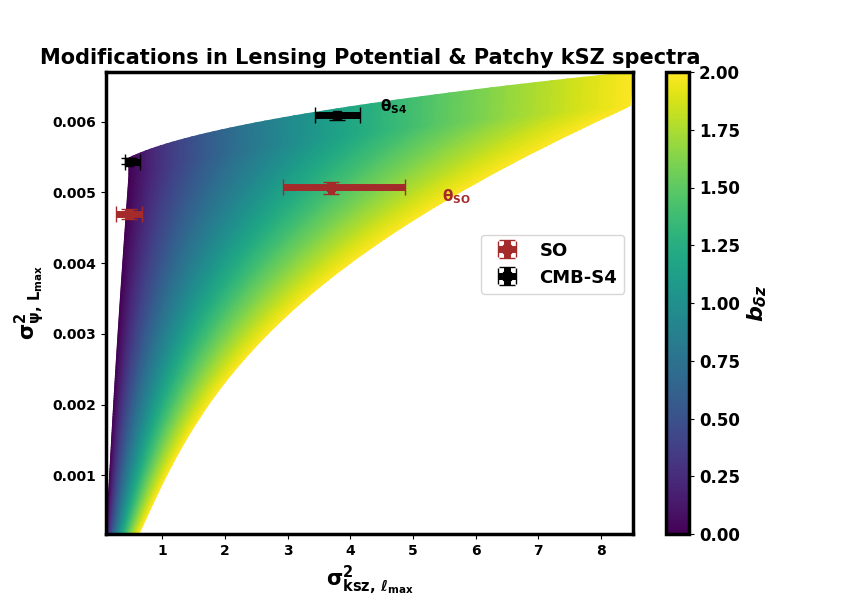}
    \caption{The patchy-kSZ and lensing potential variance (with varying $\ell_{\rm{max}}$ and $L_{\rm{max}}$) for $\Lambda$CDM $(b_{\delta z} = 0)$ and different modified $\Lambda$CDM cases using parameter $(b_{\delta z} > 0)$. The error bars from SO and CMB-S4 for the corresponding angular resolutions are shown for the cases of $\Lambda$CDM $(b_{\delta z} =0 )$ and the modified cosmology from JWST fit $(b_{\delta z} =1.15 )$. Here, the minimum multipole for lensing power spectra is taken as $L_{\rm{min}} = 50$ and for patchy-kSZ is taken as $\ell_{\rm{min}} = 2500$. }
    \label{fig:colorp}
\end{figure}

The electron density bias $b_e$ is constrained by kSZ observations, while both kSZ and CMB lensing constrain modifications to the matter power spectrum through the parameter $b_{\delta z}$. These two parameters are anti-correlated in kSZ analyses, but their degeneracy can be broken using CMB lensing, which depends only on the matter distribution.

{As has been described in Sec. \ref{sec:jwst} that the JWST observations can be explained using modified cosmology, we check the abilities of upcoming CMB surveys such as SO and CMB-S4 in being able to probe these small-scale modifications, with a joint analysis involving CMB-lensing and patchy-kSZ estimation. Thus, we consider the cases of unmodified $\Lambda$CDM (with modification  $b_{\delta z} = 0$), and best-fit modified-$\Lambda$CDM case (with $b_{\delta z} = 1.15$), obtained in Sec. \ref{sec:jwst}. We perform a Bayesian estimation of parameters $b_{\delta0}$, $b_{\delta z}$ and $b_{e} \times 10^{7}$ with emcee \citep{foreman2013emcee}, the details of which are described in Appendix \ref{sec:bayes}. 
 The JWST best-fit  power spectra at various frequencies are obtained from the input modified CMB-lensing and patchy-kSZ power spectra, with  Gaussian random noise from the respective estimators. The model power spectra are obtained by introducing modifications to $\Lambda$CDM power spectrum.}

 The corner plot of the posteriers obtained for the JWST best-fit cosmology is shown in Fig. \ref{fig:bayes}.
 {As shown in Fig. \ref{fig:bayes}, only-lensing estimation (left) is able to constrain cosmological modifications, while lensing + kSZ (right) can constrain modifications in the understanding of galactic astrophysics as well.  The cosmological  modification parameter $b_{\delta z}$ is  constrained by both the kSZ and CMB lensing observations, while the modified astrophysical scenarios are constrained via the electron density bias parameter $b_e$.  These two parameters are anti-correlated in kSZ analyses, as kSZ spectrum depends on the combination of the two using Eq. \eqref{eq:pqsum} . But their degeneracy can be broken using CMB lensing, which depends only on the matter distribution. This will allow us to put constraints on the variation of the electron density bias with multipoles and rule out modified astrophysical scenarios. Hence, a joint CMB lensing + kSZ analysis provides a robust method to probe the high-redshift structure formation, the understanding of which is being challenged by JWST observations.}

For fiducial $\Lambda$CDM case, $b_{\delta z}$ can be constrained to $< 0.26$ (95\% C.I.) with SO and $< 0.14$ with CMB-S4. For the JWST best-fit cosmology ($b_{\delta z} = 1.15$), a joint lensing and kSZ analysis can constrain modifications to $\Lambda$CDM at high redshifts ($z > 6$) and small scales ($k > 0.1\,\rm{Mpc}^{-1}$). SO can detect deviations from $\Lambda$CDM at $10.4\sigma$, and CMB-S4 at $29.8\sigma$, with CMB-S4 providing tighter constraints due to higher sensitivity. {In contrast, an only-lensing analysis can detect the deviations from $\Lambda$CDM at $8.56 \sigma$ and $23.5 \sigma$ with SO and CMB-S4 respectively.} The CMB lensing-patchy-kSZ combination thus offers smoking-gun signatures of small-scale power enhancements and potential new physics. Future experiments such as CMB-HD \citep{sehgal2019cmb} will further extend sensitivity to smaller scales.  
We use modified power spectra to compute covariances, yielding conservative parameter constraints, which would be tighter if fiducial spectra are used. Also, improved foreground modeling or template matching \citep{bennett2003first,hinshaw2007three} will strengthen constraints, whereas poorly modeled residuals may bias them.  

{There is an inherent scale and redshift dependence of electron density bias $b_e$ that results in variation across the CMB multipoles and represents an average measure of the astrophysical contribution to the integrated kSZ signal. Its value depends on stellar emission and subsequent IGM ionization, making it difficult to model due to our ignorance of the reionization process. As a result, we have used the electron density bias model which is based on semi-analytical simulations and agrees with the kSZ measurements from SPT data \citep{shaw2012deconstructing,calabrese2014precision,choudhury2021cosmic}.   
As a result, the fiducial model considered in the analysis for the Bayesian inference is with cosmological modifications and for the bias value of fiducial astrophysical scenarios.
Also, we have considered the effect of stellar emission in Sec. \ref{sec:jwst} (Eq. \eqref{eq:abparams}) where we vary the effective parameters (denoted as $A$ and $B$) to change the shape of the UV luminosity function at various redshifts to match the JWST observations. Also, we have described the relation between the UV luminosity function and the ionization fraction in Sec. \ref{sec:kszonly} (Eqs. \eqref{eq:xe} \& \eqref{eq:dniondt}), where an increase in the UV luminosity and patchiness will increase the patchy-kSZ signal. Since the kSZ power spectrum depends on both the cosmological modification and  electron density bias, our work also intends to disentangle the two effects through CMB lensing, which will only depend on the cosmological modification. Since the electron density bias is not very well constrained and uncertain, this analysis will allow us to measure the integrated electron density bias as a function of various multipoles. Constraints higher than the fiducial value from simulations would indicate modifying the underlying astrophysics. With better constraints on the electron density bias with multipoles, we will be able to rule out various astrophysical models through their impact on the luminosity function as well as the patchy-kSZ signal.
}

The root-mean-square (RMS) variance of the lensing potential and patchy-kSZ signal (see Fig. \ref{fig:colorp}) is computed as  
\begin{equation}
\sigma_{X,\ell}^2 = \sum_{\ell_{\min}(L_{\min})}^{\ell_{\max}(L_{\max})}\frac{(2\ell + 1)C^X_{\ell}}{4\pi},
\end{equation}
with $\ell_{\min} = 2500$, $L_{\min} = 50$, and $X \in \{\Psi, \rm{kSZ}\}$. The lensing RMS increases with $b_{\delta z}$ due to amplified small-scale power, while the kSZ RMS depends on both $b_{\delta z}$ and $b_e$. Enhanced lensing at $z > 6$ and $k > 0.1\,\rm{Mpc}^{-1}$ can probe $b_{\delta z} > 0.3$ using $\ell > 800$ multipoles in SO and CMB-S4.  We also show the error bars obtained using SO and CMB-S4 for the cases of $\Lambda$CDM cosmology ($b_{\delta z} = 0$) and the modified cosmology from JWST fit ($b_{\delta z} = 1.15$). 
The root-mean-square (RMS) of the  lensing potential for the cases of unmodified $\Lambda$CDM and modified $\Lambda$CDM with SO are obtained as: $4.69 _{-0.07}^{+0.07} \times 10^{-3}$ ($\Lambda$CDM) and $5.07_{-0.1}^{+0.07} \times 10^{-3}$ (modified-$\Lambda$CDM). The corresponding values with CMB-S4 are obtained as: $5.42 _{-0.03}^{+0.06}  \times 10^{-3}$ ($\Lambda$CDM) and $6.09 _{-0.07}^{+0.02}  \times 10^{-3}$ (modified-$\Lambda$CDM). The RMS on kSZ for the cases of unmodified $\Lambda$CDM and modified $\Lambda$CDM with SO are obtained as: $0.462 _{-0.199}^{+0.221}$ ($\Lambda$CDM) and $3.697 _{-0.779}^{+1.177}$ (modified-$\Lambda$CDM). The corresponding values with CMB-S4 are obtained as: $0.466 _{-0.064}^{+0.179}$ ($\Lambda$CDM) and $3.786 _{-0.344}^{+0.369}$ (modified-$\Lambda$CDM).

The lensing potential RMS increases with $b_{\delta z}$ due to enhanced small-scale power, while the patchy-kSZ RMS also rises but depends on the electron-density bias $b_e$, which can change the slopes for different cosmologies. Enhanced lensing from high redshifts ($z > 6$) at scales $k > 0.1\,\mathrm{Mpc}^{-1}$ can probe modifications $b_{\delta z} > 0.3$ using multipoles $\ell > 800$ with SO and CMB-S4. If no such enhancement is observed, the JWST excess likely stems from modified stellar astrophysics. A joint CMB-lensing and patchy-kSZ analysis will also disentangle cosmological and astrophysical effects for modified cosmology, since lensing constrains $b_{\delta z}$ independently of $b_e$, breaking the degeneracy between the two (see Table \ref{tab:kszlensall}).

\section{Conclusion}

Observations of bright galaxies by JWST have challenged the $\Lambda$CDM cosmology and our understanding of stellar dynamics. The excess UV luminosity can arise either from modified $\Lambda$CDM cosmology or changes in stellar formation and emission physics. In this work, we propose a new technique using CMB secondary anisotropies—specifically lensing and kSZ—to distinguish between these possibilities and identify whether cosmological or astrophysical processes drive the high-redshift JWST observations.  

Upcoming CMB experiments can test whether the observed excess results from modified astrophysics or small-scale ($k > 0.1\,\mathrm{Mpc}^{-1}$) cosmological changes at high redshifts ($z > 6$) via joint CMB lensing and kSZ analyses. For the JWST best-fit modified-$\Lambda$CDM case ($b_{\delta z}=1.15$), SO can constrain $b_{\delta z}$ at $10.4\sigma$ from fiducial $\Lambda$CDM, while CMB-S4 improves this to $29.8\sigma$. This approach also decouples cosmological and astrophysical effects on the patchy-kSZ signal through the parameters $b_{\delta z}$ and $b_e$ (see Fig.~\ref{fig:flow}). Future high-resolution surveys such as CMB-HD will further tighten these constraints.  

{Certain astrophysical effects such as baryonic feedback will affect CMB-lensing signal \citep{mccarthy2022baryonic,upadhye2024non}.  But these effects affect the lensing potential at very small scales ($k>$ 1 $\rm{Mpc^{-1}}$) and their impact will vary with redshifts, depending on the galaxy evolution. The effect is generally a suppression of the lensing potential at high multipoles. This effect can be analyzed by studying the distribution of clusters across redshifts, using X-rays or SZ effect \citep{birkinshaw1999sunyaev, rosati2002evolution}. Also, using cross CMB-galaxy lensing can help in ascertaining the impact of this contribution at various scales and redshifts \citep{hand2015first,singh2017cross}. }

Probing such modifications through CMB observables not only explains JWST findings but also refines our understanding of early-universe astrophysics and cosmology. Improved models of early star formation, feedback, and dust will test whether standard $\Lambda$CDM can reproduce the abundance and spectra of early galaxies \citep{vogelsberger2020high,zavala2023dusty,lovell2023extreme,Chakraborty:2025cbs}. If not, small-scale phenomenological extensions \citep{Joyce:2014kja,DelPopolo:2016emo,Tulin:2017ara,Bullock:2017xww,Sailer:2021yzm,Perivolaropoulos:2021jda,menci2022high,Abdalla:2022yfr} may be required. Furthermore, joint JWST, CMB lensing, and patchy-kSZ studies can constrain the electron-density bias $b_e$, revealing the evolution of UV luminosities and ionized bubble morphologies—offering crucial insight into the reionization epoch. Such a high-redshift analysis will also address the low-redshift related cosmological tensions \citep{baldi2016structure,cuceu2023constraints,bargiacchi2023tensions}. Thus, deeper and wider JWST surveys in combination with CMB lensing and patchy-kSZ provides a robust technique allowing us to answer the uncertainty around high redshift observations by providing important insights into the astrophysical and cosmological effects from the reionization epoch.

\section*{ACKNOWLEDGEMENT}
    This work is a part of the $\langle \texttt{data|theory}\rangle$ \texttt{Universe-Lab}, supported by the TIFR  and the Department of Atomic Energy, Government of India. The authors express their gratitude to the $\langle \texttt{data|theory}\rangle$ \texttt{Universe-Lab} and the TIFR CCHPC facility for meeting the computational needs. 
 Also, the following packages were used for this work: Astropy \citep{astropy:2013,astropy:2022,astropy:2018},
, NumPy \citep{harris2020array}
CAMB \citep{2011ascl.soft02026L}, SciPy \citep{2020SciPy-NMeth}, SymPy \citep{10.7717/peerj-cs.103}, Matplotlib \citep{Hunter:2007}, emcee \citep{foreman2013emcee}, HEALPix (Hierarchical Equal Area isoLatitude Pixelation of a sphere)\footnote{Link to the HEALPix website http://healpix.sf.net}\citep{2005ApJ...622..759G,Zonca2019}, PySM \citep{Thorne_2017} and Cluster Toolkit \citep{2022ascl.soft09004M}.

\bibliographystyle{aasjournalv7}

\bibliography{references}{}

\appendix

\section{Bayesian Estimation of the JWST best-fit cosmology}
\label{sec:bayes}

{As has been shown in Fig. \ref{fig:dndm}, that the JWST observations can be explained using modified cosmology, we check the abilities of upcoming CMB surveys such as SO and CMB-S4 in being able to probe these small-scale modifications, with a joint analysis involving CMB-lensing and patchy-kSZ estimation. Thus, we consider the cases of unmodified $\Lambda$CDM (with the modification  $b_{\delta z} = 0$), and the best-fit modified-$\Lambda$CDM case (with $b_{\delta z} = 1.15$) obtained in Sec. \ref{sec:jwstsummary}. We show the results for the beams and noises corresponding to the frequencies 93 (beam= 2.2 arcmin, $\Delta_T = 8 \, \mu$K-arcmin ) and 145 GHz (beam= 1.4 arcmin, $\Delta_T = 10 \, \mu$K-arcmin ) for SO-baseline \citep{SimonsObservatory:2018koc,Ade_2019} configuration, and 95 (beam= 2.2 arcmin, $\Delta_T = 2.9 \, \mu$K-arcmin ) and 145 GHz (beam= 1.4 arcmin, $\Delta_T = 2.8 \, \mu$K-arcmin ) for CMB-S4 \citep{abazajian2016cmbs4}. The polarization noise is taken to be $\Delta_P = \sqrt{2} \Delta_{T}$ for both the configurations.}

We perform a Bayesian estimation of the parameters $b_{\delta0}$, $b_{\delta z}$ and $b_{e} \times 10^{7}$ , with emcee \citep{foreman2013emcee} using the Markov Chain Monte Carlo approach \citep{heavens2009statistical,hobson2010bayesian,verde2010statistical,trotta2017bayesian}. 
We use a flat prior from 0 to 10 for all parameters ($b_{\delta0}$, $b_{\delta z}$ and $b_{e} \times 10^{7}$) and a Gaussian likelihood. 
The lensing estimation is able to provide constraints on the modifications to the power spectrum. 
The posterior distribution obtained from lensing estimation is then used as prior for kSZ estimation. The log likelihood for both estimations is given as:
\begin{equation}
-2\log \mathcal{L}_X = \sum_{\nu} \sum_{\ell = \ell_{\rm{min}} (L_{\min})}^{\ell = \ell_{\rm{max}}(L_{\max})} \left[ \frac{(  C_{\ell, X}^{\nu}|_{\rm{data}} - C_{\ell, X}^{\nu}|_{\rm{model}})^2}{\,  \rm{Cov_{\ell, X}^{\nu}}} + \log{(2\pi \,  \rm{Cov_{\ell, X}^{\nu}}) } \right],
\end{equation}
{here $X \in \{\Psi , \rm{kSZ}\}$, and we use $\ell_{\rm{min}} = 2500$ and $\ell_{\rm{max}}$ corresponding to the beam size for the various frequency bands, indicated by $\nu$. Also, we use $L_{\min} = 50$. For SO, we use $L_{\max} = 1500$, while for  CMB-S4, we use $L_{\max} = 2000$. The JWST best-fit  power spectra at various frequencies $\nu$, which are given as ($C_{\ell}^{\nu}|_{\rm{data}}$), are obtained from the input modified CMB-lensing and patchy-kSZ power spectra, with  Gaussian random noise from the respective estimators. The model power spectra are obtained by introducing modifications to the $\Lambda$CDM power spectrum.}
{We perform Bayesian estimation with 20 walkers for 30000 steps. We discard 20000 steps to remove the burn-in part of the chains and perform thinning on the rest for every 20 step size. }

The posteriors obtained on the modified lensing potential and patchy-kSZ power spectrum parameters are shown in Fig. \ref{fig:bayes}. The injected values from the modified cosmology model that agrees with the JWST high-redshift observations, are shown in blue lines (with $b_{\delta 0} = 1$, $b_{\delta z} = 1.15$ and $b_{e} = 4 \times 10^{-7}$). The corner plot shows that the parameters $b_{\delta z}$ and $b_e$ that affect the patchy-kSZ power spectrum are anticorrelated. This degeneracy is broken using CMB lensing observations.



\end{document}